\documentclass[english,aps,floats,twocolumn,showpacs,nofootinbib]{revtex4}
\usepackage{pslatex}
\usepackage[T1]{fontenc}
\usepackage[latin1]{inputenc}
\usepackage{graphicx}
\usepackage{epsfig}

\usepackage{calc}
\usepackage{ifthen}

{
{
{
\newcommand{\bea}{\begin{eqnarray}}
\newcommand{\eea}{\end{eqnarray}}

\newcommand{\nc}{\newcommand}
\nc{\renc}{\renewcommand}
\nc{\eqs}[2]{\mbox{Eqs.~(\ref{#1},\,\ref{#2})}}
\nc{\eq}[1]{\mbox{Eq.~(\ref{#1})}}
\nc{\figs}[2]{\mbox{Figs.~(\ref{#1},\,\ref{#2})}}
\nc{\fig}[1]{\mbox{Fig~.(\ref{#1})}}
\nc{\be}[1]{\begin{equation} \mbox{$\label{#1}$}}
\nc{\ee}{\vspace{0.1cm}\end{equation}}

\newcommand{\bean}{\begin{eqnarray*}}
\newcommand{\eean}{\end{eqnarray*}}

\def\rhob{\rho_{b\;s}}

\begin{document}
\title{Baryon-to-Dark Matter Ratio from Random Angular Fields}
\author{John McDonald}
\email{j.mcdonald@lancaster.ac.uk}
\affiliation{Lancaster-Manchester-Sheffield Consortium for Fundamental Physics, Cosmology and Astroparticle Physics Group, Dept. of Physics, University of 
Lancaster, Lancaster LA1 4YB, UK}
\begin{abstract}

     We consider the baryon-to-dark matter ratio in models where the dark matter and baryon densities depend on angular fields $\theta_{d}$ and $\theta_{b}$ according to $\rho_{d} \propto \theta_{d}^{\alpha}$ and $\rho_{b} \propto \theta_{b}^{\beta}$, with all values of $\theta_{d}$ and $\theta_{b}$ being equally probable in a given randomly-selected domain. Under the assumption that anthropic selection depends primarily on the baryon density in galaxies at spherical collapse, we show that the probability density function for the baryon-to-dark matter ratio $r = \Omega_{B}/\Omega_{DM}$ is purely statistical in nature and is independent of  anthropic selection. We compute the probability density function for $r$ as a function of $\alpha$ and $\beta$ and show that the observed value of the baryon-to-dark matter ratio, $r \approx 1/5$, is natural in this framework.

\end{abstract}
% \pacs{}
\maketitle

\section{Introduction}

      The ratio of the baryon and dark matter mass densities  is observed to be $\Omega_{B}/\Omega_{DM} \approx 1/5$, where $DM$ denotes cold dark matter \cite{wmap}. A number of different strategies to have been pursued to account for this. A common origin for dark matter and baryons, with a conserved quantum number directly linking their number densities, leads to asymmetric dark matter with dark matter particle mass is in the range 1-10 GeV. Alternatively, the dark matter and baryon density could be determined by physically similar processes. For example, a weak-strength annihilation process might separately account for both the baryon asymmetry and thermal relic WIMP dark matter densities \cite{bm1,bm2,wimpy}.

      All of these approaches account for the baryon-to-dark matter ratio via a physical mechanism which directly  determines the ratio as a function of particle physics model parameters. The alternative is anthropic selection. This relies on the existence of superhorizon-sized domains with varying $\Omega_{B}/\Omega_{DM}$. One way this can be achieved is if the baryon or dark matter densities are determined by a scalar field which is massless during inflation. An example based on axion dark matter was suggested in \cite{lindeax}, while a varying baryon density and the associated baryon isocurvature constraints were explored generally in \cite{aadb1}, with a particular example being Affleck-Dine baryogenesis. The possibility of superhorizon-sized domains of varying baryon number was first proposed in \cite{lindenew}, in the context of Affleck-Dine baryogenesis, where it was also noted that a range of baryon density at spherical collapse was likely to be anthropically preferred. 

  However, simply varying one or other of the densities cannot by itself completely explain why $\Omega_{B}$ is similar to $\Omega_{DM}$. For example, in \cite{lindeax} the dark matter density varies but the baryon density is assumed to be equal to that in the observed Universe. The argument of \cite{lindeax} focuses on how increasing the dark matter density will affect the baryon and dark matter densities in galaxies, resulting in conditions which are likely to disfavour the existence of observers. This can explain why $\Omega_{B}/\Omega_{DM} \sim 1$ only if we can understand why the observed baryon density is close to some critical density above which the evolution of observers is disfavoured. But how did the observed baryon density in our domain get close to this critical density in the first place? To explain this we need to let both the baryon and dark matter densities vary between domains. 

     Here we consider a plausible class of models for the generation of domains of varying baryon and dark matter densities. We will assume that both the baryon and dark matter densities are determined by separate angular variables, derived from fields which are effectively massless during inflation. A specific example is a combination of axion dark matter \cite{lindeax} and Affleck-Dine baryogenesis \cite{aadb1,lindenew}. In the following we will study this class of model completely generally. The anthropic selection criterion we will apply is that there is an anthropically preferred baryon density in galaxies at spherical collapse, such that the probability of the evolution of observers is insensitive to the baryon-to-dark matter ratio.  This is consistent with the idea that baryon densities in galaxies which are significantly larger than that in the observed Universe are anthropically disfavoured \cite{lindeax,lindenew}. 

    The paper is organized as follows. In Section 2 we introduce the model and establish the baryon and dark matter densities in galaxies at spherical collapse. In Section 3 we derive the probability density function for the baryon-to-dark matter ratio in domains which have equal baryon density at spherical collapse. We then apply this function to the case of axion dark matter combined with a baryon asymmetry determined by a random CP phase ({\it Anthropic Baryogenesis} \cite{aadb1}) and to other plausible models. In Section 4 we discuss the plausibility of the anthropic selection criterion we have used. In Section 5 we summarize our conclusions.

\section{Dark matter and baryon densities due to random angular fields}  

  The class of model we consider is based on the assumption that the dark matter and baryon densities are given by 
\be{n1}  \rho_{dm} \propto \theta_{d}^{\alpha} \;\;\;;\;\;\rho_{b} \propto \theta_{b}^{\beta}     ~\ee
where $\theta_{d}$ and $\theta_{b}$ are angular fields which are assumed to have equal probablility for all values of the angles. The best known example of this is axion dark matter. In realistic models the behaviour will change 
as the angles approach $\pi$, but we will consider the cases of interest to have $\theta_{d}, \; \theta_{b} \ll 1$. This is expected in anthropic selection models, since the anthropically selected domains usually have atypically small values of the random angles \cite{lindeax,aadb1}.

     Due to either random initial conditions at the onset of inflation or quantum fluctuations of the angular fields, superhorizon domains of essentially constant $\theta_{d}$ and $\theta_{b}$ are naturally generated \cite{aadb1}. Therefore a horizon-sized Universe at the present time will have a random value of $(\theta_{d},\theta_{b})$, with all such pairs of values having equal probability.

      We next consider the growth of perturbations in domains with different baryon and dark matter densities. We assume that only parameters which vary between the domains are the baryon and dark matter densities.  We will keep all other parameters constant, in particular the magnitude of the primordial density perturbation. 

     In the following it will be important to distinguish between mean baryon and dark matter densities in a given domain at a given temperature (denoted by $\rho_{b}$ and $\rho_{dm}$) and the mean densities is galaxy-sized perturbations at spherical collapse (denoted $\rhob$ and $\rho_{dm\;s}$). 
  We first consider the effect on galaxies of varying $(\theta_{d},\theta_{b})$. Varying $(\theta_{d},\theta_{b})$ will change the mean dark matter and baryon densities in the domain at a given $T$. Increasing the mean density will lead to a greatly enhanced density in galaxy-sized perturbations at spherical collapse. This is because matter-radiation equality then occurs at a higher temperature. As a result, significant perturbation growth (growth proportional to the scale factor), which begins approximately at matter-radiation equality, will begin at a higher temperature and so the perturbation will become non-linear and undergo spherical collapse at a correspondingly higher temperature and density, enhancing the baryon and dark matter densities in galaxies. 

    We next derive the enhancement factors for the baryon and dark matter densities in galaxies at spherical collapse. Let the mean baryon and dark matter densities 
at a given $T$ in a reference domain be $\rho_{b\;o}$ and 
$\rho_{dm\;o}$.  (This could be our domain, but we will keep it general.)  Suppose in a different domain these are enhanced by factors $f_{b}$ and $f_{d}$, such that $\rho_{b} = f_{b} \rho_{b\;o}$ and $\rho_{dm} = f_{d} \rho_{dm\;o}$. The total density at $T$ in this domain is then 
\be{n2} \rho_{TOT} = f_{d}\rho_{dm\;o} + f_{b} \rho_{b\;o}   ~.\ee
Therefore 
\be{n3} \gamma \equiv \frac{\rho_{TOT}}{\rho_{TOT\;o}} = \frac{(f_{d} + f_{b} r_{o})}{(1 + r_{o}) }     ~,\ee 
where $r_{o}$ is the baryon-to-dark matter ratio in the reference domain and we have defined $\gamma$.  
The matter-radiation equality temperature is determined by 
\be{n4}  K T_{eq}^{4} = \rho_{TOT}(T_{eq})   ~,\ee
where $K = (\pi^2 g(T)/90 M_{p}^2)^{1/2}$ and $g(T)$ is the effective number of massless degrees of freedom. 
Thus 
\be{n5} \frac{T_{eq}^{4}}{T_{eq\;o}^{4}}  =  \frac{\rho_{TOT}(T_{eq})}{\rho_{TOT}(T_{eq_{o}})}   ~.\ee      
Using 
\be{n6}  \rho_{TOT}(T_{eq\;o}) = \rho_{TOT}(T_{eq}) \frac{T_{eq_{o}}^{3}}{T_{eq}^{3}}  ~,\ee
we obtain
\be{n7}  \frac{T_{eq}}{T_{eq\;o}} = \frac{f_{d} + f_{b} r_{o}}{1 + r_{o}} \equiv \gamma  ~.\ee
Assuming that the galaxy-sized perturbations enter the horizon during radiation-domination (this will generally be satisfied except at very small values of $r$), the temperature at which spherical collapse occurs will also be enhanced by $\gamma$, since the rate of growth of the perturbations is the same once the universe becomes matter-dominated. The baryon and dark matter densities at spherical collapse will then be enhanced by a factor $\gamma^{3}$ from the earlier time of spherical collapse and by factors $f_{d}$ and $f_{b}$ from the enhancement of the dark matter and baryon density at a fixed temperature. Therefore 
\be{n8} \frac{\rho_{dm\;s}}{\rho_{dm\;s_{o}}} = \gamma^{3} f_{d} \;\;\;;\;\; 
\frac{\rho_{b\;s}}{\rho_{b\;s_{o}}} = \gamma^{3} f_{b}   ~,\ee
where $\rho_{b\;s}$ and $\rho_{dm\;s}$ are the baryon and dark matter densities in galaxies at spherical collapse\footnote{These expressions are equivalent to the more complicated expression given in \cite{aadb1}.}.

   The anthropic selection criterion we will apply in the following is that the probability of the evolution of observers is dependent only on the baryon density at spherical collapse and is independent of the dark matter density in galaxies and so of $r$. This is not likely to be exactly true, but we will argue later that it should be a good approximation to a realistic anthropic selection criterion. We will see that the effect of this condition is that the probability of $r$ factors out of the total probability and becomes purely statistical in nature, allowing us to determine its probability density exactly.

   For a given value of $\rhob$, the value of $f_{d}$ is constrained to be a function of $f_{b}$, 
\be{n9} f_{d} = \left(\frac{\rho_{b\;s}}{\rho_{b\;s_{o}}}\right)^{1/3} 
\frac{\left(1 + r_{o}\right)}{f_{b}^{1/3}} - f_{b} r_{o}  ~.\ee 
This can also be written in terms of $\theta_{d}$ and $\theta_{b}$, using $f_{d} = (\theta_{d}/\theta_{d\;o})^{\alpha}$ and 
$f_{b} = (\theta_{b}/\theta_{b\;o})^{\beta}$. For simplicity, in the following we will absorb $\theta_{d\;o}$ and $\theta_{b\;o}$ into the definition of $\theta_{d}$ and $\theta_{b}$, so that 
$f_{d} = \theta_{d}^{\alpha}$ and $f_{b} = \theta_{b}^{\beta}$, 
with $\theta_{d\;o} = \theta_{b\;o} = 1$. Then the constraint is 
\be{n9a} \theta_{d} = \left(
\left(\frac{\rho_{b\;s}}{\rho_{b\;s_{o}}}\right)^{1/3} 
\frac{\left(1 + r_{o}\right)}{\theta_{b}^{\beta/3}} - r_{o} \theta_{b}^{\beta}  \right)^{\frac{1}{\alpha}}    
~.\ee
The physical interpretation of this constraint is that there are a range of different mean baryon and dark matter densities (and so baryon-to-dark matter ratios) which can lead to the same baryon density at spherical collapse. For example, reducing the mean dark matter density causes the matter-radiation equality temperature to decrease, reducing the baryon density at spherical collapse. This can be compensated by increasing the mean baryon density.

\section{The probability density function for the baryon-to-dark matter ratio}

    We next calculate the probability density function for $r$ in domains satisfying the constraint that the baryon density at spherical collapse is between $\rho_{b\;s_{o}}$ and $\rho_{b\;s_{o}} +  \Delta \rho_{b\;s_{o}}$, where $\Delta \rho_{b\;s_{o}} \ll \rho_{b\;s_{o}}$ and $\Delta \rhob$ is a small change over which the anthropic probability of $\rhob$ does not vary significantly. 

      Since the probability of each value of $\theta_{b}$ and each value of $\theta_{d}$ is equal, the probability of finding the pair $(\theta_{b},\theta_{d})$ in a given domain to be within a range of values corresponding to an area of the $(\theta_{b},\theta_{d})$ plane is simply proportional to the area. In Figure 1 we plot the constraint \eq{n9a} for $\rho_{b\;s}$ equal to $\rho_{b\;s_{o}}$ and for $\rho_{b\;s}$ equal to $\rho_{b\;s_{o}} + \Delta  \rho_{b\;s_{o}}$, for the case $(\alpha,\beta) = (2,1)$. The total probability for $\rho_{b\;s}$ to be in the range $\rho_{b\;s_{o}}$ to $\rho_{b\;s_{o}} + \Delta  \rho_{b\;s_{o}}$ is then proportional to the area between these curves. 

   To obtain the probability density with respect to $r$, we need to find the area between the curves for a small change $\Delta r$. In Figure 2 we show the change in $(\theta_{b},\theta_{d})$ from a point on the first curve due to (i) a change in $r$ for fixed $\rho_{b\;s}$ and (ii) a change in $\rho_{b\;s}$ for fixed $r$. The corresponding area is then the area of the parallelogram with sides described by the vectors ${\bf v}_{1}$ and ${\bf v}_{2}$, 
\be{n13} {\rm Area} = |{\bf v}_{1} {\bf \times} {\bf v}_{2}| = |\Delta \theta_{d} \delta \theta_{b} - \Delta \theta_{b} \delta \theta_{d} | ~\ee
where (with $i = b$ or $d$)
\be{n14} \Delta \theta_{i} = \frac{\partial \theta_{i}}{\partial \rho_{b\;s}} \Delta  \rho_{b\;s_{o}}  ~\ee
and  
\be{n15} \delta \theta_{i} = \frac{\partial \theta_{i}}{\partial r} \Delta r   ~.\ee

 To evaluate the derivatives, we impose the relation between $\theta_{d}$, $\theta_{b}$ and $r$, 
\be{n16} \theta_{d} = \left(\frac{r_{o}}{r}\right)^{\frac{1}{\alpha}} \theta_{b}^{\frac{\beta}{\alpha}}   ~\ee
and the constraint \eq{n9a} 
to obtain $\theta_{b}$ and $\theta_{d}$ as functions of $r$ and $\rhob$ 
\be{n17} \theta_{b} = \left( \frac{A \rhob^{1/3}}{r_{o}} 
\left(\frac{r}{1 + r}\right) \right)^{\frac{3}{4 \beta}} ~\ee  
and
\be{n18} \theta_{d} = 
\left(\frac{A \rho_{b\;s}^{1/3}}{\theta_{b}^{\beta/3}} - r_{o} \theta_{b}^{\beta}  \right)^{\frac{1}{\alpha}}    
~,\ee
where we have defined $A = (1 + r_{o})/\rho_{b\;s_{o}}$ and we have kept the dependence on $r$ and $\rhob$ explicit. The derivatives as a function of $r$ and $\rhob$ are then
\be{d1}   
\frac{\partial \theta_{b}}{\partial r} = 
\frac{3}{4 \beta} \left( \frac{A \rhob^{1/3}}{r_{o}}\right)^{\frac{3}{4 \beta}}
\frac{1}{\left(1 + r\right)^2}  
\left(\frac{r}{1 + r}\right)^{\frac{3}{4\beta} -1} 
~,\ee
\be{d2}   
\frac{\partial \theta_{d}}{\partial r} = 
\frac{-r_{o}^{\frac{1}{\alpha}}}{4 \alpha} 
\left( \frac{A \rhob^{1/3}}{r_{o}}\right)^{\frac{3}{4 \alpha}}
\frac{1}{r^{\frac{1}{\alpha} +1} } 
\left(1 + \frac{3r}{1 + r} \right) 
\left(\frac{r}{1 + r}\right)^{\frac{3}{4\alpha}} 
~,\ee
\be{d3}  \frac{\partial \theta_{b}}{\partial \rho_{b\;s}} =
\frac{1}{4 \beta \rho_{b\;s}} 
\left( \frac{A \rhob^{1/3}}{r_{o}} 
\left(\frac{r}{1 + r}\right) \right)^{\frac{3}{4 \beta}}
~,\ee
and
\be{d4}   
\frac{\partial \theta_{d}}{\partial \rho_{b\;s} } = 
\frac{1}{4 \alpha \rho_{b\;s}} 
\left( \frac{r_{o}}{r} 
\left(\frac{A \rhob^{1/3}}{r_{o}}\right)^{3/4} 
\left(\frac{r}{1 + r}\right)^{3/4} \right)^{\frac{1}{\alpha}} 
~.\ee  
The area is then 
\be{d5}  K  
 \left(\frac{r}{1 + r}\right)^{\frac{3}{4}\left(\frac{1}{\alpha} + \frac{1}{\beta}\right)} 
\frac{1}{r^{\frac{1}{\alpha} + 1}} \Delta \rho_{b\;s} 
\Delta r   ~\ee
where 
\be{d6} K = \frac{r_{o}^{\frac{1}{\alpha}} }{4 \alpha \beta \rho_{b\;s}}  \left(\frac{1 + r_{o}}{r_{o}}\right)^{\frac{3}{4}\left(\frac{1}{\alpha} + \frac{1}{\beta}\right)} 
~.\ee
Although the first curve is defined to have value $\rho_{b\;s_{o}}$, we have now dropped the subscript and consider the probability density to be a function of $\rhob$.  
The terms $K$ and $\Delta \rho_{b\;s}$ are independent of $r$ and so can be absorbed into a normalization for the probability density as a function of $r$.
Thus the probability density for a domain with a fixed value of $\rhob$ to have a given value of the baryon-to-dark matter ratio $r$ is
\be{d7}  f(r) = N \left(\frac{r}{1 + r}\right)^{\frac{3}{4}\left(\frac{1}{\alpha} + \frac{1}{\beta}\right)} 
\frac{1}{r^{\frac{1}{\alpha} + 1}}  ~,\ee
where $N$ is a normalization constant which depends on $\alpha$ and $\beta$. 

    The integral of \eq{d7} from $r = 0$ to $\infty$ is convergent provided that 
$\alpha > 0$ (convergent as the upper limit tends to $\infty$) and $\alpha > \beta/3$
(convergent as the lower limit tends to zero). These conditions are likely to be satisfied for 
realistic models, in which case the probability density is well-defined.

 From \eq{d5} we see that the probability density for $\rhob$ and $r$ factors into an $r$ dependent factor and a $\rho_{b\;s}$ dependent factor. The total probability density of finding a given value of $\rho_{b\;s}$ and $r$ is then obtained by  
multiplying the probability density in \eq{d5} by an anthropic selection factor which depends only on $\rhob$.

   The interpretation of the probability density \eq{d7} is that it gives the probability for any observer with a given value of $\rhob$ to find themselves in a domain with a given value of $r$. The probability of $r$ is purely statistical, since $r$ is assumed to have no anthropic effect on the evolution of observers with a given value of $\rhob$, and simply represents the relative number of domains with $\rhob$ which have a given value of $r$, all of which are equally likely for the observer to have evolved in. This is justified in so far as the dark matter density in galaxies does not have a strong effect on the evolution of observers i.e. the baryon density at spherical collapse is the dominant anthropic parameter.

\subsection{Application to the axion-anthropic baryogenesis model} 

    As a first example, we will apply the probability density \eq{d7} to the case of axion dark matter combined with an anthropic baryogenesis model, for example Affleck-Dine baryogenesis \cite{aadb1,lindenew}. In this case $\rho_{d} \propto \theta_{a}^{2}$ and $\rho_{b} \propto  \theta_{b}$, corresponding to $(\alpha,\beta) = (2,1)$. The probability density is then 
\be{n24} f(r) = \frac{N}{r^{3/8} \left(1 + r\right)^{9/8} }  ~,\ee
where the normalization factor is given by  
\be{n26} N = \frac{\Gamma(\frac{9}{8})}{\sqrt{\pi} \Gamma(\frac{5}{8}) } \approx 0.370   ~.\ee
In Figure 3 we plot the probability density per logarithmic interval, $r \times f(r)$, which gives an indication of the likelihood of finding $r$ close to a given value. ($f(r)$ itself is divergent as $r \rightarrow 0$.) The maximum is at $r = 1.25$. 

   In Figure 4 we show the probability of $r$ being within ranges which differ by factors of 10. The corresponding probabilities are given in Table 1.  We find that 30.3$\%$ of domains have $r$ in the range 0.1-1. Thus it is natural in this model for the baryon-to-dark matter ratio to have the observed value $r = 0.2$. However, it is not 
the case that values similar to 1 are inevitable. From Figure 4 and Table 1 we see that 13.5$\%$ of domains have $r < 0.1$, with 3.2$\%$ having $r < 0.01$, while 21.8$\%$ of domains have $r > 10$, with 9.5$\%$ having $r > 50$. Thus there is also a significant probability that a given observer will evolve in a galaxy which is strongly dominated by dark matter or in a galaxy with almost no dark matter. 

  Nevertheless, we find that 65$\%$ of domains have $r$ in the range 0.1-10. Therefore models of this type are clearly able to explain why the observed
density of baryons is within an order of magnitude of the observed density of dark matter.

    The probability density favours lower values of $r$ within a given range. This can be seen in Table 2, where we show the range $r = 0.1-1.1$ divided into intervals with $\Delta r = 0.2$. This shows that domains with $r$ around 0.2 are favoured out of the subset of domains with $r = 0.1 - 1.1$.

\begin{figure}[htbp]
\begin{center}
\epsfig{file=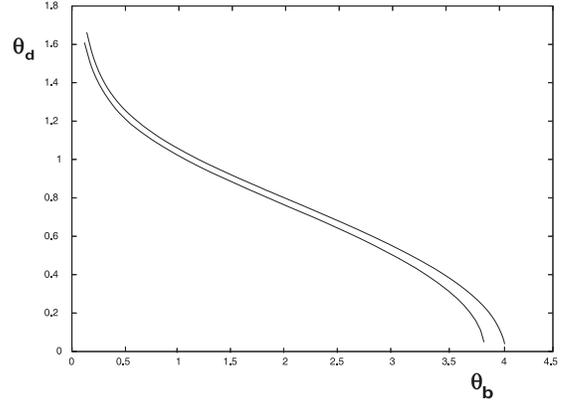, width=0.3\textwidth, angle = -90}
\caption{Region of the $(\theta_{b},\theta_{d})$ plane satisfying the constraint of equal baryon density at spherical collapse}
\label{fig2}
\end{center}
\end{figure}

\begin{figure}[htbp]
\begin{center}
\epsfig{file=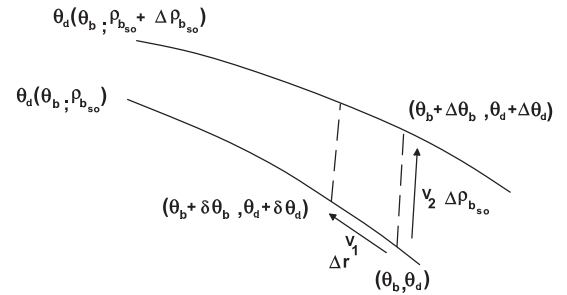, width=0.4\textwidth, angle = -0}
\caption{Calculation of probability density function.}
\label{fig2}
\end{center}
\end{figure}

\begin{figure}[htbp]
\begin{center}
\epsfig{file=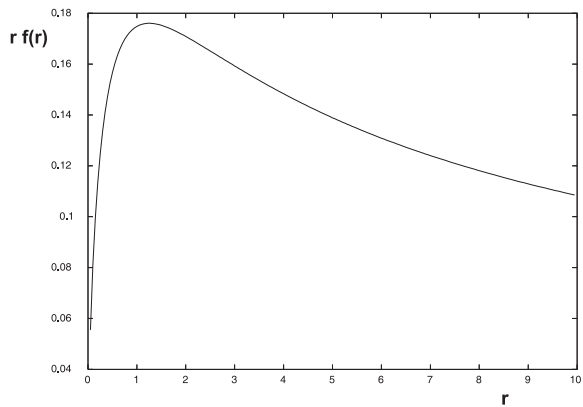, width=0.3\textwidth, angle = -90}
\caption{Probability density function per logarithmic interval as a function of $r$ for the case $(\alpha,\beta) = (2,1)$.}
\label{fig2}
\end{center}
\end{figure}

\begin{figure}[htbp]
\begin{center}
\epsfig{file=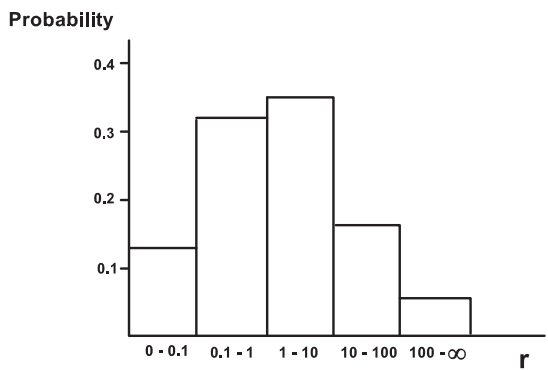, width=0.4\textwidth, angle = -0}
\caption{Probability of the baryon-to-dark matter ratio for the axion-anthropic baryogenesis model, $(\alpha,\beta) = (2,1)$.}
\label{fig2}
\end{center}
\end{figure}

\subsection{Other models}

   We show in Table 3 the probabilities for three other plausible values of $\alpha$ and $\beta$:  (i) $(\alpha,\beta) = (1,1)$, (ii)  $(\alpha,\beta) = (1,2)$
and (iii)  $(\alpha,\beta) = (2,2)$. We see that all of these models can naturally account for the observed value of $r$, with the fraction of domains in the range $r = 0.1-1$ being 
40.9$\%$, 24.9$\%$ and 47.6$\%$ for models (i), (ii) and (iii) respectively. Therefore the ability of this class of model to account for the observed baryon-to-dark ratio is not strongly dependent upon the values of $\alpha$ and $\beta$. 
Model (ii) has a much larger fraction of domains with a large dark matter content, corresponding to $r < 0.1$. This can be understood as due to the stronger dependence of $\theta_{d}$ on $\rho_{d}$ ($\theta_{d} \propto \rho_{d}$) and weaker  dependence of $\theta_{b}$ on $\rho_{b}$ ($\theta_{b} \propto \rho_{b}^{1/2}$). This means that a greater increase in the probability of a domain is obtained by increasing $\rho_{d}$ by a factor than by increasing $\rho_{b}$ by the same factor.  

\begin{table}[h]
\begin{center}
\begin{tabular}{|c|c|}
 \hline $r$	&	$p$  \\
\hline	$0-0.1 $	&	$0.135$		\\   	
\hline	$0.1-1 $	&	$0.303$  	\\   		
\hline	$1-10 $	&	$0.344$		\\ 
  \hline	$10-100 $	&	$0.154$		\\   	
\hline	$100-\infty $	&	$0.064$  	\\   		
\hline\hline	$0-0.01 $	&	$0.032$		\\  
\hline	$10-50 $	&	$0.123$		\\   
\hline	$50-\infty $	&	$0.095$		\\   		
\hline     
 \end{tabular} 
 \caption{\footnotesize{Probability of $r$ for the axion-anthropic baryogenesis model 
($\alpha = 2$, $\beta = 1$). }}  
 \end{center}
 \end{table}

\begin{table}[h]
\begin{center}
\begin{tabular}{|c|c|}
 \hline $r$	&	$p$  \\
\hline	$0.1-0.3 $	&	$0.115$		\\   	
\hline	$0.3-0.5 $	&	$0.073$  	\\   		
\hline	$0.5- 0.7$	&	$0.054$		\\ 
  \hline	$0.7-0.9$	&	$0.042$		\\   	
\hline	$0.9-1.1$	&	$0.035$  	\\   		
\hline     
 \end{tabular} 
 \caption{\footnotesize{Probability of $r$ for the axion-anthropic baryogenesis model     
 ($\alpha = 2$, $\beta = 1$) in the range 0.1-1.1.}}  
 \end{center}
 \end{table}

\begin{table}[h]
\begin{center}
\begin{tabular}{|c|c|c|c|}
 \hline $r$	&	$p(1,1)$  &  $p(1,2)$ & $p(2,2)$\\
\hline	$0-0.1 $	&	$0.296$	& $0.633 $	&	$0.385$	 	\\   	
\hline	$0.1-1 $	&	$0.409$      & $0.249$	&	$0.476$		\\   		
\hline	$1-10 $	&	$0.248$	& $0.101 $	&	$0.214$		\\ 
  \hline	$10-100 $	&	$0.042$	&	$0.015 $	&	$0.086$	 \\   	
\hline	$100-\infty $	&	$0.005$ &  $0.002 $	&	$0.035$	 	\\   		
\hline\hline	$0-0.01$	&	$0.093$	&	$0.380 $	&	$0.189$	 \\  
\hline	$10-50 $	&	$0.037$	&	$0.013 $	&	$0.070$	 \\   
\hline	$50-\infty $	&	$0.010$ & $0.003 $	&	$0.052$		\\   \hline     
 \end{tabular} 
 \caption{\footnotesize{Probability of $r$ for alternative models with  
($\alpha$, $\beta$) $=$ (1,1), (1,2) and (2,2). }}  
 \end{center}
 \end{table}

\section{Plausibility of the Criterion for Anthropic Selection}

    We have assumed that the main parameter determining the evolution of observers is the baryon density at spherical collapse, and that the density of dark matter at spherical collapse has only a small effect. How good an approximation is this likely to be? 

    Suppose we consider the effect of (i) increasing the 
baryon density at spherical collapse relative to our own domain while keeping the
dark matter density constant and (ii) increasing the dark matter density at spherical collapse while keeping the baryon density constant.

     We first consider the effect of increasing the baryon density. It is believed that the balance between dynamical collapse and radiative cooling favours formation of galaxies with baryon mass 
$\sim 10^{12}M_{\odot}$, corresponding to the largest mass for which a collapsing gas cloud can cool, and that this mass is independent of the conditions at spherical collapse (being expressible in terms of fundamental constants) \cite{silketal}. In this case, increasing the baryon density will mean that the volume at spherical collapse which contains $\sim 10^{12}M_{\odot}$ becomes smaller, 
therefore galaxies will form from smaller length perturbations. As a result there will be more galaxies within a given volume than in our domain. (After spherical collapse, we can expect the time necessary for observers to evolve to be similar in all domains, $\sim 1-10$ Gyr.) Since stars are forming out of an initially more dense gas cloud, it is also plausible that larger and therefore shorter-lived stars will tend to form. In addition, since the density perturbation is unchanged, these smaller length perturbations, which would correspond to galaxy subhalo scales in our domain, can be expected to be highly clustered. Therefore we can expect multiple mergers of galaxies, resulting in large densities of stars and potential disruption of planet orbits. Moreover, if the baryon density becomes large enough, the size of the volume with $\sim 10^{12}M_{\odot}$ will become smaller than the size of observed galaxy discs, in which case the whole picture of galaxy formation would have to be radically altered. In particular, as noted in \cite{lindenew}, the density of stars will eventually be such that the spacing between stars is comparable to the size of a solar system, at which point stable planet orbits will become unlikely; this would represent the upper limit on the baryon density at spherical collapse.  Therefore it is plausible that increasing the baryon density at spherical collapse will have a strong effect on the probability of forming long-lived stars with stable planet orbits in a radiation-friendly environment, which is the primary requirement for the evolution of observers. This is consistent with the idea that anthropic selection is sensitive to the baryon density at spherical collapse and that there is an anthropically preferred value of the baryon density at spherical collapse. It is also consistent with the naive expectation that increasing the baryon density in the volume corresponding to our galaxy at spherical collapse will result in a correspondingly larger density of stars, which is the basis of \cite{lindeax} and \cite{aadb1}.     

   We next consider the effect of increasing the dark matter density at spherical collapse while keeping the baryon density constant. 
In this case the perturbations corresponding to baryon mass $\sim 10^{12}M_{\odot}$ will have the same size as in our domain. The only effect will be to increase the dark matter density in the galaxy halos. While this may have some effect on the subsequent evolution of the galactic disc, it would not be expected to have a catastrophic effect on the probability of forming long-lived stars with stable planet orbits, except in the limit of a very large increase in the dark matter density, corresponding to $r \rightarrow 0$. (With fixed baryon density at spherical collapse, $r$ effectively parameterizes the dark matter density at spherical collapse.) However, domains with larger dark matter densities, corresponding to $r < 0.1$, generally account for only a small fraction of the total probability for $r$. For example, in the case of axion dark matter combined with anthropic baryogenesis ($(\alpha, \beta) = (2,1)$), we find that only 13.5$\%$ of domains have $r < 0.1$. We do not expect domains with smaller dark matter densities, corresponding to $r > 0.1$, to have a very strong anthropic dependence on $r$.  Moreover, if the effect of a large dark matter density is to disfavour the evolution of observers, the effect would be to the redistribute the total probability density as a function of $r$ to larger values of $r$, which only strengthens the conclusion that $r = 0.2$ can be naturally understood in this framework. Therefore our conclusions should be robust with respect to the anthropic effects of the dark matter density.  

    There have been other approaches to determining the axion dark matter density anthropically \cite{wilczek,axias,axmit}. In particular, in \cite{axias} it was noted that decreasing the temperature at matter-radiation equality while keeping the density perturbation constant implies that galaxies will not be able to achieve spherical collapse before entering the dark energy dominated era. This imposes a lower bound on the axion density. We have not considered dark energy here, but we note that if we are to understand the coincidence of the dark energy density with the observed baryon and dark matter densities, it is likely that the dark energy density must also be domain-dependent, in which case anthropic constraints based on a fixed dark energy density do not apply. We will return to this issue in future work.

\section{Conclusions}

    We have shown that baryon and dark matter densities which depend on random angular fields, when combined with the assumption that the dominant anthropic selection parameter is the baryon density in galaxies at spherical collapse, can easily account for why the observed baryon-to-dark matter ratio is approximately  equal to 1/5. The total probability for the baryon and dark matter densities in a given domain is factorized into an anthropic factor, which depends only on $\rhob$, and a 
purely statistical factor, which depends only on $r$. 

  The model is crucially dependent upon both the baryon and dark matter densities being able to randomly vary between superhorizon-sized domains. To make this clear, we can contrast our model with the model of \cite{lindeax}, in which only the axion density varies, with the baryon density being fixed to its observed value. In this case changing the dark matter density changes the baryon density at spherical collapse. Therefore $r$ is determined anthropically, not statistically as in the model we have presented here. Moreover, in the model of \cite{lindeax}, the baryon density in our domain must be tuned to be close to some anthropically critical baryon density, so that increases in the dark matter density to values larger than we observe are anthropically disfavoured. No such tuning is required in the model presented here, since the baryon density can be naturally anthropically selected, and the value of $r$ is determined purely statistically and not anthropically.

      %%%%%%%%%%%%%%%%%%%%%%%%%%%%%%%%%%%%%%%%%%%%%%%%%%%%%%%%%%%%%%%%

\section*{Acknowledgements}
The work of JM is supported by the Lancaster-Manchester-Sheffield Consortium for Fundamental Physics under STFC grant
ST/J000418/1.

%%%%%%%%%%%%%%%%%%%%%%%%%%%%%%%%%%%%%%%%%%%%%%%%%%%%%%%%%%%%%%%%%%%%


\begin{thebibliography}{99}

\bibitem{wmap}  E.~Komatsu {\it et al.}  [WMAP Collaboration],
  %``Seven-Year Wilkinson Microwave Anisotropy Probe (WMAP) Observations: Cosmological Interpretation,''
  Astrophys.\ J.\ Suppl.\  {\bf 192}, 18 (2011)
  [arXiv:1001.4538 [astro-ph.CO]].
  %%CITATION = ARXIV:1001.4538;%%


\bibitem{bm1} J.~McDonald,
  %``Baryomorphosis: Relating the Baryon Asymmetry to the 'WIMP Miracle',''
  Phys.\ Rev.\ D {\bf 83}, 083509 (2011)
  [arXiv:1009.3227 [hep-ph]].
  %%CITATION = ARXIV:1009.3227;%%

\bibitem{bm2} J.~McDonald,
  %``Simultaneous Generation of WIMP Miracle-like Densities of Baryons and Dark Matter,''
  Phys.\ Rev.\ D {\bf 84}, 103514 (2011)
  [arXiv:1108.4653 [hep-ph]].
  %%CITATION = ARXIV:1108.4653;%%

\bibitem{wimpy} Y.~Cui, L.~Randall and B.~Shuve,
  %``A WIMPy Baryogenesis Miracle,''
  JHEP {\bf 1204}, 075 (2012)
  [arXiv:1112.2704 [hep-ph]].
  %%CITATION = ARXIV:1112.2704;%%


\bibitem{lindeax}  A.~D.~Linde,
  %``Inflation and Axion Cosmology,''
  Phys.\ Lett.\ B {\bf 201}, 437 (1988).
  %%CITATION = PHLTA,B201,437;%%


\bibitem{aadb1}  J.~McDonald, %``Anthropically Selected Baryon Number and Isocurvature Constraints,''
JCAP {\bf 1210}, 005 (2012)
[arXiv:1207.2135 [hep-ph]].
%%CITATION = ARXIV:1207.2135;%% 


\bibitem{lindenew} A.~D.~Linde, %``The New Mechanism Of Baryogenesis And The Inflationary Universe,''
Phys.\ Lett.\ B {\bf 160} (1985) 243.
%%CITATION = PHLTA,B160,243;%% 


\bibitem{silketal} J.~Silk,  \nat {\bf 265}, 710 (1977); Rees, M.~J. \& Ostriker, J.~P.,  Mon.\ Not.\ Roy.\ Astron.\ Soc.\ {\bf 179}, 541 (1977). 
 
\bibitem{wilczek} F.~Wilczek, %``A Model of anthropic reasoning, addressing the dark to ordinary matter coincidence,''
In *Carr, Bernard (ed.): Universe or multiverse* 151-162
[hep-ph/0408167].
%%CITATION = HEP-PH/0408167;%% 


\bibitem{axias} S.~Hellerman and J.~Walcher, %``Dark matter and the anthropic principle,''
Phys.\ Rev.\ D {\bf 72}, 123520 (2005)
[hep-th/0508161].
%%CITATION = HEP-TH/0508161;%% 


\bibitem{axmit} M.~Tegmark, A.~Aguirre, M.~Rees and F.~Wilczek, %``Dimensionless constants, cosmology and other dark matters,''
Phys.\ Rev.\ D {\bf 73}, 023505 (2006)
[astro-ph/0511774].
%%CITATION = ASTRO-PH/0511774;%% 




\end{thebibliography}
\end{document}